\documentclass[letter]{aa}
\usepackage{graphicx}
\usepackage{txfonts}
\usepackage[colorlinks=true, linkcolor=blue, citecolor=blue, urlcolor=blue]{hyperref}
\usepackage[utf8]{inputenc}
\usepackage{subcaption}
\usepackage{xcolor}
\usepackage{placeins}

\defcitealias{Soler_2025}{S25} 

\begin{document}

    \title{Are supernovae driving turbulence in the solar neighborhood?}
   \author{
        Georges Abboudeh\inst{1}
        \and
        Patrick Hennebelle\inst{1}
        \and
        Juan D. Soler\inst{2}
        \and
        Noé Brucy\inst{3}
        \and 
        Tine Colman\inst{3}
        \and
        Ralf Klessen\inst{4,5}
        \and
        Marc-Antoine Miville-Deschênes\inst{6}
        \and
        Sergio Molinari\inst{7}
        \and
        Alice Nucara\inst{7,8}
        \and 
        Veli-Matti Pelkonen\inst{7}
        \and
        Alessio Traficante\inst{7}
        \and Robin Tress\inst{9}
       }

   \institute{
        Universit\'e Paris-Saclay, Universit\'e Paris Cit\'e, CEA, CNRS, AIM, F-91191 Gif-sur-Yvette, France
        \and
        University of Vienna, Department of Astrophysics, T\"urkenschanzstra\ss e~17, 1180 Wien, Austria
        \and
        ENS de Lyon, CRAL UMR5574, Universite Claude Bernard Lyon 1, CNRS, Lyon 69007, France
        \and
        Universit\"at Heidelberg, Zentrum f\"ur Astronomie, Institut f\"ur Theoretische Astrophysik, Albert-Ueberle-Str.~2, 69120 Heidelberg, Germany
        \and
        Universit\"at Heidelberg, Interdisziplin\"ares Zentrum f\"ur Wissenschaftliches Rechnen, Im Neuenheimer Feld~225, 69120 Heidelberg, Germany
        \and
        Laboratoire de Physique de l’Ecole Normale Supérieure, ENS, Université PSL, CNRS, Sorbonne Université, Université Paris Cité, Observatoire de Paris, Paris, France
        \and
        INAF - Istituto di Astrofisica e Planetologia Spaziali, Via Fosso del Cavaliere 100, I-00133 Roma, Italy
        \and
         Dipartimento di Fisica, Università di Roma Tor Vergata, Via della Ricerca Scientifica 1, 00133 Roma, Italy
         \and
         Institute of Physics, Laboratory for Galaxy Evolution and Spectral Modelling, EPFL, Observatoire de Sauverny, Chemin Pegasi 51, 1290 Versoix, Switzerland
        }

   \date{Received 21 January 2026  /
Accepted 12 March 2026 }


  \abstract
   {Turbulence plays an important role in shaping the interstellar medium, and it strongly influences star formation.}
   {We aim to identify the physical processes capable of sustaining \ion{H}{i} turbulence in the solar neighborhood. }
   {We compare recent \ion{H}{i} line-of-sight velocity observations within a
    volume of radius 70--500~pc centered on the Sun with a suite of $1$~kpc numerical simulations that include two distinct turbulent drivers: (i) supernova (SN) feedback and (ii) imposed large-scale turbulent forcing. For each simulation, we constructed synthetic sky maps that closely mimic the observational one, allowing for a consistent comparison between the simulations and the observational data.}
   {The \ion{H}{i} observations show a median velocity dispersion of $11.1~\mathrm{km\,s^{-1}}$ in the solar neighborhood. The SN-driven simulations systematically underpredict this value, yielding dispersions in the range $4.9$–$6.7~\mathrm{km\,s^{-1}}$. We find that the simulations with strong enough large-scale forcing can reproduce not only the median observed velocity dispersion but also the observed velocity distribution.}
   {}

   \keywords{Turbulence -- solar neighborhood --local insterstellar matter-- ISM: kinematics and dynamics
               }

\maketitle

\section{Introduction}
The interstellar medium (ISM) is the birthplace of stars, and to better understand star formation, it is essential to study the mechanisms that occur in the ISM.
Turbulence is one of the main regulators of star formation, alongside gravity \citep{federrath_2012}. 
It plays a dual role, either inducing or opposing gravitational collapse \citep{mac_low_2004}.

Many processes contribute to turbulence in the ISM. However, their relative contributions differ, and their importance varies across the different scales \citep{hennebelle_2012, klessen_2016}.
 Several studies \citep[e.g.,][]{Wada_2002, Bournaud_2010, Fensch_2023} have shown that large-scale gravitational instabilities, driven by self-gravity and galactic rotation, can sustain ISM turbulence even in the absence of stellar feedback. 
Other large-scale processes, such as gas accretion into the galaxy, have also been found to be capable of driving turbulence in the ISM \citep{klessen_2010}. 
 Among the different feedback processes, supernova (SN) explosions from massive stars are believed to dominate the energy injection into the ISM \citep{klessen_2016}. Each event releases $\sim10^{51}\,\mathrm{erg}$ in a blast wave, which injects both thermal energy and radial momentum into the surrounding gas. The exact contribution from large driving and SNe remains controversial and may depend on the environment \citep{iffrig_2015, Walch_2015, Ohlin_2019}. 

\citet{krumholz_2018} argue that in galaxies with low gas surface densities, stellar feedback alone can drive turbulence, whereas in gas-rich high-redshift disks, turbulence is mainly sustained by energy injection from large-scale gravitational inflows. 
A similar result was obtained by \citet{brucy_2020, brucy_2023} using 1~kpc ISM-box simulations. They showed that in gas-rich galaxies, the SN alone cannot sustain enough turbulence to maintain the star formation rate close to observed levels.
Observations from \citet{bacchini_2020} showed that in nearby galaxies with low surface density disks, SNe can sustain turbulence without the need for external large-scale drivers. 
\citet{Roman-Oliveira_2024} explored gas-rich galaxies at a redshift of $z\approx4.5$ and found that the measured velocity dispersions were consistent with the feedback-dominated branch of the \citet{krumholz_2018} model. However, they found that including gravitational instabilities overestimated the turbulence.
 
On smaller scales, simulations of an SN-driven 250-pc-side ISM box by \citet{padoan_2016} showed that SN explosions can drive and sustain turbulence within molecular clouds. By comparing their SN-driven cloud sample with observations from the Outer Galaxy Survey, they obtained consistent cloud statistics and dynamical properties. \citet{colman_2022} compared 1~kpc ISM-box simulations with Herschel observations of the Large Magellanic Cloud and found a similar result on cloud scales. However, on scales larger than $\sim$60~pc, SN feedback alone is not sufficient, and an additional large-scale driver is required.

We note that \ion{H}{i} 21-cm observations have long been used to study the kinematics and turbulence of the neutral ISM in nearby galaxies \citep{McClure-Griffiths_2023}. 
Large surveys such as THINGS \citep{Walter_2008} have provided high-resolution maps of \ion{H}{i} across the disks of nearby galaxies. Using these data, \citet{Tamburro_2009} found systematically higher velocity dispersions ($\sigma_{\mathrm{\ion{H}{i}}} \simeq 10 \pm 2~\mathrm{km\,s^{-1}}$) in the inner actively star-forming regions of galaxies and lower dispersions ($\sigma_{\mathrm{\ion{H}{i}}} \simeq 6~\mathrm{km\,s^{-1}}$) in the outer \ion{H}{i} disks. These results suggest that stellar feedback drives the turbulence in star-forming disks, while large-scale galactic processes maintain a baseline level of turbulence at larger radii.

In the nearby ISM, \cite{Soler_2025} (from here on \citetalias{Soler_2025}) combined three-dimensional dust models from \citet{Edenhofer_2024} with \(\mathrm{H\,\textsc{i}}\;21\text{ cm}\) and \(\mathrm{CO}(1\!-\!0)\) emission to construct a spatial map of the line-of-sight (LOS) velocity \((v_{\mathrm{los}}\)) for atomic and molecular gas within a region spanning
$70\text{--}1250~\mathrm{pc}$ in distance and $|b|<5^{\circ}$ around the Sun.
From these data, they calculated the mean velocity dispersions in $\mathrm{H\,\textsc{i}}$ and \(\mathrm{CO}\):
\(\sigma_{v,\mathrm{H\,\textsc{i}}}=10.8\,\mathrm{km\,s^{-1}}\) and \(\sigma_{v,\mathrm{CO}}=6.6\,\mathrm{km\,s^{-1}}\), respectively.

In this work, we compare the observed dispersions of \ion{H}{i} \(v_{\mathrm{los}}\)  with numerical simulations in which turbulence is driven either by (i) SN feedback or (ii) stochastic Fourier-space forcing, which is meant to represent turbulent energy injected from the large scales. We also include a comparison with the fiducial \texttt{MHD-4pc} run from the TIGRESS suite \citep{tigress_2017}, where turbulence is also driven by SN feedback. Our goal is to determine which mechanism can reproduce the observed level of turbulence in the local ISM.

The paper is organized as follows. Section 2 presents a brief description of the numerical simulations used. In Sect. 3, we describe the analysis pipeline and present the velocity dispersion comparison. In Sect. 4, we summarize our findings and evaluate whether SN feedback or large-scale driving can reproduce the observed level of turbulence in the local ISM.

\section{Numerical setup}\label{numerical_setup}

\begin{table*}[ht]
\caption{Simulation parameters.} 
\centering
\begin{tabular}{lcccccccc}
\hline\hline
Simulation
& $f_{\rm rms}$
& $\Sigma$
& $B_{0}$
& Coarse resolution
& Finest resolution
& Time
& $\sigma_{v,\mathrm{los}}$ \\
& 
& (M$_{\odot}$\,pc$^{-2}$)
& ($\mu$G)
& (pc)
& (pc)
& (Myr)
& (km\,s$^{-1}$) \\
\hline
TURB-7     & 7000  & 19.1  & 3.5 & 4 & 0.5  & 229--237 & 9.61\\
TURB-10    & 10000 & 19.1  & 3.5 & 4 & 0.5  & 215--225 & 13.36\\
TURB-15    & 15000 & 19.1 & 3.5 & 4 & 0.5  & 189--199 & 12.75\\
TURB-20    & 20000 & 19.1  & 3.5 & 4 & 0.5 & 153--156 & 15.27\\
\hline
SN-0.66    & 0     & 8.4  & 2.1 & 4 & 1    & 65--104  & 4.94\\
SN-1.0     & 0     & 12.7 & 2.1 & 4 & 1    & 42--50   & 5.37\\
SN-1.5     & 0     & 19.1 & 2.1 & 4 & 1    & 42--60   & 6.07\\
TIGRESS-MHD4pc & 0 & 13 & 2.6 & 4 & 4    & 200--400 & 6.65\\
\hline
\end{tabular}

\tablefoot{
The \texttt{TURB} runs correspond to simulations with imposed large-scale turbulent forcing,
and \texttt{SN} runs correspond to SN-driven simulations \citep{Colman_2025}. 
We note that \texttt{TIGRESS-MHD4pc} is also SN-driven \citep{tigress_2017}.  
The term $f_{\rm rms}$ sets the amplitude of the imposed large-scale turbulent forcing and is expressed in units of $1.46\times10^{-4}\,\mathrm{km\,s^{-1}\,Myr^{-1}}$.
The term $\Sigma$ is the initial gas column density, and $B_{0}$ is the initial magnetic-field strength.
"Coarse resolution" and "Finest resolution" give the coarsest and finest
grid spacings.
"Time" indicates the interval over which the synthetic maps were
constructed. 
The term $\sigma_{v,\mathrm{los}}$ is the mean LOS velocity dispersion computed from the synthetic maps of each simulation after applying a LOS velocity threshold of $\lvert v_{\mathrm{los}} \rvert \le 30~\mathrm{km\,s^{-1}}$.
}
 
\label{tab:sim_params}
\end{table*}

In this section we briefly present the numerical simulations used for comparison with the \ion{H}{i} data. The simulations fall into two main families that differ in their mechanism for driving turbulence. Table~\ref{tab:sim_params} provides a summary of the main simulation parameters. Section~\ref{frigg_section} introduces the \texttt{FRIGG} simulations, which include both SN-driven and large-scale–forced runs, while Sect.~\ref{tigress_section} presents the SN-driven \texttt{MHD-4pc} simulation from the TIGRESS suite \citep{tigress_2017}.

\subsection{FRIGG simulations}\label{frigg_section}

The \texttt{FRIGG} simulations consist of a suite of three-dimensional simulations of a cubic $1~\mathrm{kpc}$ region of the ISM. 
The simulations include two distinct families that differ mainly in their turbulence driving mechanism: SN-driven runs (\texttt{SN}), taken from
\citet{Colman_2025}, and large-scale turbulent-forcing runs (\texttt{TURB}), introduced in this work. 

These simulations follow a setup similar to that of the \texttt{FRIGG} simulation presented by \citet{Hennebelle_2018} and were performed using the adaptive mesh refinement code \texttt{RAMSES} \citep{teyssier_2002}, which solves the equations of ideal magnetohydrodynamics (MHD; \citealt{fromang_2006}). The base grid consists of $256^{3}$ cells, with periodic boundary conditions in the disk plane ($x$ and $y$) and open boundaries in the vertical ($z$) direction. 

Each simulation begins with an initial turbulent velocity field generated with random phases and a Kolmogorov power spectrum.
Turbulence is then sustained either by SN feedback (\texttt{SN} runs) or by stochastic large-scale forcing applied in Fourier space (\texttt{TURB} runs). 
Details of the turbulence driving prescriptions are provided in
Appendix~\ref{app:driving}.

\subsection{TIGRESS simulation}\label{tigress_section}
Our reference model is the \texttt{MHD-4pc} run of the TIGRESS suite \citep{tigress_2017}.  
It follows a local shearing box that spans \(L_x=L_y=1024\;\mathrm{pc}\) in the disk plane and \(L_z=4096\;\mathrm{pc}\) vertically.
The grid is uniform, with a 4 pc cell size. The boundary conditions are shearing-periodic in the \(x\) direction, periodic in \(y\), and vacuum in \(z\).
The simulation starts with an initial \(k^{-2}\) velocity field for \(t = 50\;\mathrm{Myr}\) and then decay it to zero by \(t = 100\;\mathrm{Myr}\) to prevent an artificial collapse burst.
After that, turbulence is driven self‐consistently by clustered and
runaway SNe together with time‐dependent far-ultraviolet heating; no
external forcing is applied.
The simulation reaches a statistical steady state after
\(\sim 100\;\mathrm{Myr}\) and is evolved for a total of
\(700\;\mathrm{Myr}\). More information about this simulation can be found in their paper.
For our analysis, we used the publicly available snapshots,\footnote{\url{https://princetonuniversity.github.io/astro-tigress/}}
which consist of a \(1024^{3}\;\mathrm{pc^{3}}\) sub‐volume centered on the
mid-plane and cover the interval
\(t \simeq 200\text{–}400\;\mathrm{Myr}\).

\section{Results}

\begin{figure}[t]
  \centering
  \includegraphics[width=\linewidth]{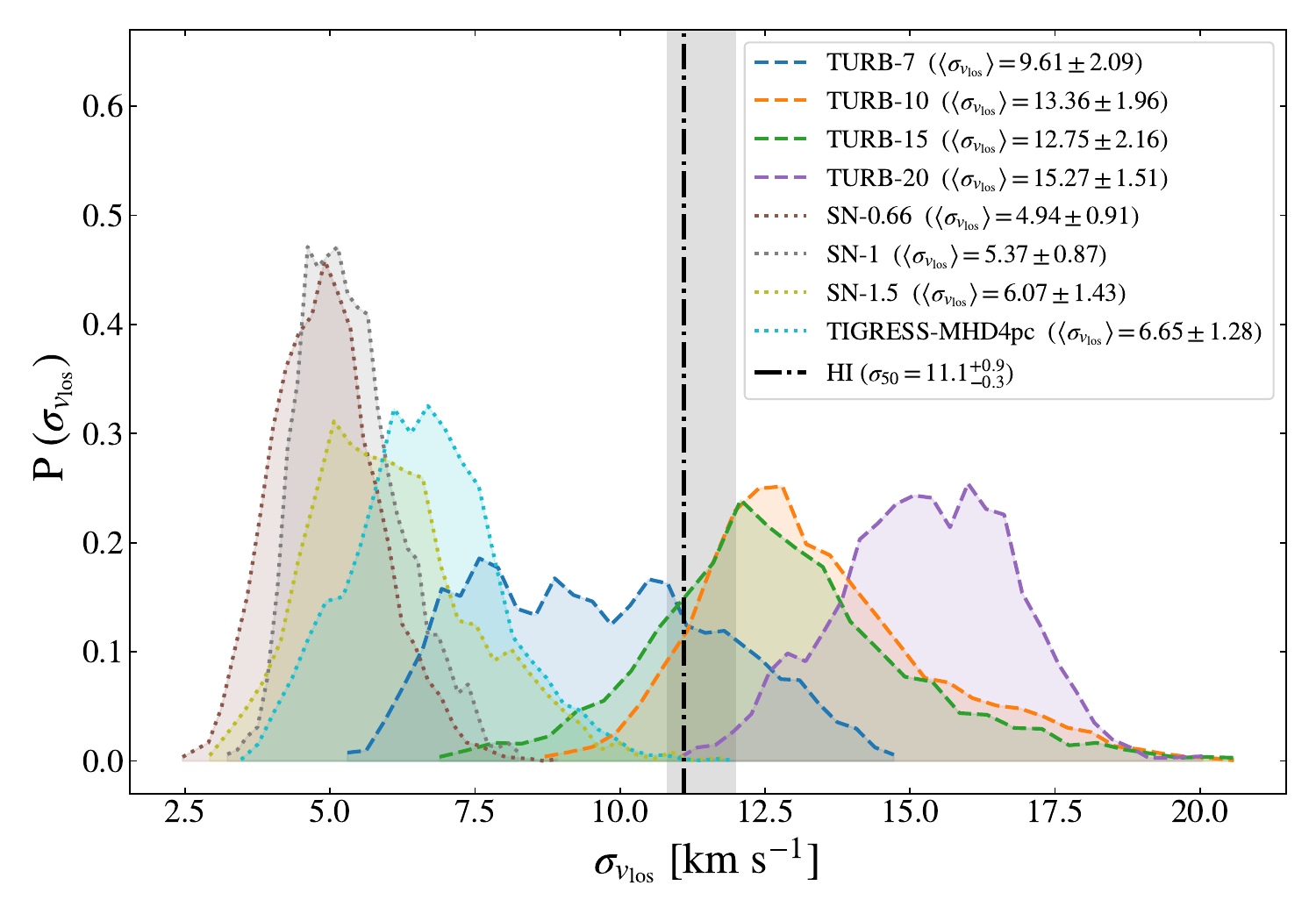}
 \caption{Probability density functions of mass-weighted LOS velocity dispersions for all simulations. Each curve shows the distribution of dispersion values obtained from the different masked sky maps for a given simulation. Dashed curves correspond to the large-scale–forced \texttt{TURB} runs, while dotted curves denote the SN-driven simulations (\texttt{SN} series and \texttt{TIGRESS-MHD4pc}). 
The dot-dashed vertical line marks the median \ion{H}{i} velocity dispersion, $\sigma_{50} = 11.1^{+0.9}_{-0.3}~\mathrm{km\,s^{-1}}$, derived from the Monte Carlo sampling of the LOS velocities reported by \citetalias{Soler_2025} using the \citet{Reid_2019} Galactic rotation model parameters in a volume spanning 70--500~pc around the Sun. The gray shaded vertical band represents the corresponding 25th--75th percentile range. 
Velocity dispersions were computed after applying a velocity threshold of $|v_{\mathrm{los}}| \le 30~\mathrm{km\,s^{-1}}$.}
\label{fig:sigma_distribution}
\end{figure}

\subsection{The dispersions measured by observation}

\citetalias{Soler_2025} measured LOS \ion{H}{i} velocities in the solar neighborhood by combining three-dimensional dust density models from \citet{Edenhofer_2024} with \ion{H}{i} line emission. This approach allowed them to reconstruct the local \ion{H}{i} velocity field. However, the initial reconstruction includes velocity components that may not be physically associated with the local ISM. Instead, they may arise from chance correlations in the method or from nonlocal high-velocity gas along the LOS. 

Motivated by the LOS velocities expected in the solar neighborhood from the Galactic rotation model of \citet{Reid_2019}, \citetalias{Soler_2025} imposed a velocity cut of $|v_{\mathrm{los}}| \le 25~\mathrm{km\,s^{-1}}$ on the input \ion{H}{i} spectra to mitigate spurious correlations and exclude nonlocal emission. They then subtracted the Galactic rotation model from the reconstructed velocities. In the following, when referring to observational LOS velocities, we always subtract the Galactic rotation model.
After subtracting it, the resulting LOS velocities can exceed the original \(|v_{\mathrm{los}}| \le 25~\mathrm{km\,s^{-1}}\) cut applied to the input \ion{H}{i} spectra. 

We restricted the comparison of the simulations to the distance range
\(70\text{--}500~\mathrm{pc}\). In this range, the maximum LOS velocity after subtracting the model is \(\sim 30~\mathrm{km\,s^{-1}}\). We therefore adopted \(|v_{\mathrm{los}}| \le 30~\mathrm{km\,s^{-1}}\) as the velocity threshold for comparison with the simulations. This choice is further supported by the fact that up to this threshold, the velocity dispersions measured in the inner and outer Galaxy remain comparable, whereas they increasingly diverge at higher velocity thresholds (see Appendix~\ref{app:diff_vel_thresholds}).

\subsection{Line-of-sight velocity construction}\label{sec:vlos_def}

To ensure a consistent resolution across all simulations, we first
resampled the simulation outputs onto a regular grid with a resolution
of 4~pc before constructing the LOS velocity maps. Following \citetalias{Soler_2025}, we defined the LOS velocity
seen by an observer as the projection of the local gas velocity \(\mathbf v=(v_x,v_y,v_z)\) onto the sight-line vector:

\begin{equation}
  v_{\mathrm{los}}(\mathbf r) \;=\;
  \frac{x\,v_x+y\,v_y+z\,v_z}
       {r}\;.
  \label{eq:vlos}
\end{equation}
Here, \(x,y,z\) are the coordinates
of the cell relative to the observer, and \(r = \sqrt{x^{2} + y^{2} + z^{2}}\).

\paragraph{FRIGG boxes.}
To minimize the biases that can arise from the choice of a particular position, we placed the observer successively at a set of uniformly spaced positions in the disk mid-plane.
Specifically, the observer coordinates were chosen as \((x_{\rm obs}, y_{\rm obs}, 0)\), with \(x_{\rm obs}\) and \(y_{\rm obs}\) taking values between \(-400\) and \(500~\mathrm{pc}\), sampled at intervals of \(100~\mathrm{pc}\). 
Because the boxes are strictly periodic in \((x,y,0)\), all cell coordinates are rewrapped relative to each observer.

\paragraph{TIGRESS \texttt{MHD--4pc}.}
The shearing box is periodic only in \(y\).
We therefore shifted the observer along the \(y\)-axis in steps of \(50~\mathrm{pc}\), spanning the range \(-450 \le y_{\rm obs} \le 500~\mathrm{pc}\), and along the vertical direction in steps of \(25~\mathrm{pc}\) within \(-25 \le z_{\rm obs} \le +25~\mathrm{pc}\) to increase the statistical sample. In contrast, \(x_{\rm obs}\) was kept fixed to avoid complications associated with shear-periodic boundary conditions.
No periodic wrapping was applied in \(z\). The restriction on latitude \(b\) (see below) ensures that all sight lines remain close to the mid-plane and do not cross the vertical vacuum boundaries.\\

We only kept cells with gas temperatures in the range \(100~\mathrm{K} < T < 10000~\mathrm{K}\) and distances \(70 \le r \le 500~\mathrm{pc}\). The dependence on the adopted temperature thresholds is examined in Appendix~\ref{app:diff_cuts_temp}.
The lower distance limit corresponds to the minimum distance covered by the \ion{H}{i} data.
After applying these spatial cuts, the remaining cells were binned into \(4~\mathrm{pc}\) radial bins to match the resolution of the numerical simulations.
Positions were then transformed into Galactic coordinates \((\ell,b)\), and only those with \(|b|\le5^{\circ}\) were kept. 
Finally, the sky was divided into \(10^{\circ}\!\times\!10^{\circ}\) longitude-latitude tiles, and mass-weighted averages of the LOS velocity and density were computed for each radial bin and tile.

The \ion{H}{i} map has a finer radial resolution (logarithmic bins of \(0.4\text{-}2.8~\mathrm{pc}\)) and incomplete sky coverage, unlike the simulations. 
For a consistent comparison at a similar resolution and sampling, we mapped the \ion{H}{i} data onto the same \((r,\ell)\) grid used for the simulations, with linear radial bins spanning \(70\) to \(500~\mathrm{pc}\) of \(\simeq 4~\mathrm{pc}\) and longitude bins of \(10^{\circ}\). 
Within each \((r,\ell)\) cell, we computed the mean and standard deviation of the LOS velocity, and we recorded cells with no data as "\(\mathrm{NaN}\)."
We then applied the observational \((r,\ell)\) sampling pattern to the simulation maps (produced from the different observer positions at different time steps) so that only the regions corresponding to observed cells are kept.
To minimize any possible biases caused by the nonuniform longitudinal sampling of the observations, we repeated this procedure after rotating the observational grid by \(0^{\circ}\), \(90^{\circ}\), \(180^{\circ}\), and \(270^{\circ}\) in \(\ell\), thus producing four masked realizations for each simulation map. In Appendix~\ref{app:cake_plots} we show the LOS velocity map of \ion{H}{i} from \citetalias{Soler_2025} together with an example of a full simulation map and the four masked versions obtained after applying the different observational samplings.

\subsection{Velocity dispersion comparison }\label{sec:vel_disp_comp}

After constructing the synthetic sky maps for the different simulations, a velocity cut of \(|v_{\mathrm{los}}| \le 30~\mathrm{km\,s^{-1}}\) was applied for a consistent comparison with observations. We then computed the mass-weighted LOS velocity dispersion for each map. Figure~\ref{fig:sigma_distribution} shows the obtained distribution of these values. 
The dot-dashed vertical line in the figure corresponds to the median observed \ion{H}{i} velocity dispersion, $\sigma_{50} = 11.1^{+0.9}_{-0.3}~\mathrm{km\,s^{-1}}$, derived in Appendix~\ref{app:galacticrotation} from Monte Carlo sampling of LOS velocities reported by \citetalias{Soler_2025} using 1000 realizations of the \citet{Reid_2019} Galactic rotation model (see Appendix~\ref{app:galacticrotation}).

The SN-driven simulations produced systematically lower velocity dispersions, with mean values of \( \sim 4.9\text{--}6.7~\mathrm{km\,s^{-1}} \), indicating that SN feedback alone underestimates the turbulent level observed in the solar neighborhood. Among the SN-driven simulations, the \texttt{SN-1.5} and \texttt{TIGRESS-MHD4pc} runs show the largest spreads, with tails reaching and occasionally exceeding the observational value, suggesting that rare SN-driven configurations can attain the required amplitudes but that the typical state remains under-turbulent relative to the data. 

As for the large-scale–forced runs, they tend to exhibit higher velocity dispersions at larger forcing amplitudes, with values ranging between $\sim 9.6$ and $15.3~\mathrm{km\,s^{-1}}$.
The velocity dispersion in \texttt{TURB-20} exceeds the observational reference, which lies between the values measured for \texttt{TURB-7} and the intermediate forcing runs \texttt{TURB-10} and \texttt{TURB-15}. The similar dispersions obtained for these two runs (with \texttt{TURB-10} slightly higher) despite their different forcing amplitudes are a consequence of the imposed velocity cut, which limits the contribution from the high-velocity tail of the distributions. 

Overall, SN-driven turbulence remains systematically weaker than observed. Although SN feedback may dominate turbulence on smaller scales, our results indicate that a large-scale driver is required to reproduce the observed velocity dispersion within the \(70\text{--}500\)~pc range considered here.

\section{Summary and conclusions}\label{conclusion}

We have compared solar-neighborhood \ion{H}{i} LOS velocities from \citetalias{Soler_2025} over the distance range of \(70\text{--}500\)~pc with two families of 1~kpc box simulations that differ in their turbulence drivers: (i) SN-driven runs, where turbulence is powered solely by SNe, and (ii) large-scale turbulent-forcing runs, where energy is injected at low Fourier modes to mimic the injection from the galactic turbulent cascade. Our comparison of LOS velocity dispersions showed that the SN-driven simulations systematically underpredict the observed \ion{H}{i} dispersion. The typical values obtained in the SN runs fall well below the observational range, $\sigma_{\mathrm{obs}} = 10.8\text{--}12~\mathrm{km\,s^{-1}}$, inferred from the \ion{H}{i} data. 

Assuming standard turbulent dissipation, the energy injection rate required to sustain a velocity dispersion, $\sigma$, scales as $\dot{E}\propto\sigma^{3}$. This implies that reproducing the observed \ion{H}{i} velocity dispersion requires an energy input approximately four to ten times larger than that provided by SN-driven models, and potentially even more if velocities above the adopted $ 30~\mathrm{km\,s^{-1}}$ $|v_{\mathrm{los}}|$ threshold are taken into account.
Additional energy injection on large scales is therefore required.
In our setup, we find that a large-scale forcing amplitude between those of
\texttt{TURB-7} and intermediate forcing runs (\texttt{TURB-10} and \texttt{TURB-15}) is sufficient to reproduce the observed
velocity dispersion.

We stress that these estimates should be regarded as lower limits since the adopted velocity cut is likely conservative and including higher-velocity gas would further increase the measured dispersion.
However, identifying the physical nature of the required large-scale driving will require galactic simulations that include self-consistent galactic-scale physics.

\begin{acknowledgements} 
This research has received funding from the European Research Council synergy grant ECOGAL (Grant 855130). This work was granted access to HPC resources of CINES and CCRT under the allocation x2020047023 made by GENCI (Grand Equipement National de Calcul Intensif). JDS is funded by the Austrian Science Fund (FWF) through the Principal Investigator Project Grant DOI 10.55776/PAT6169824. NB acknowledges support from the ANR BRIDGES grant (ANR-23-CE31-0005). TC acknowledges funding received from the European High Performance Computing Joint Undertaking (JU) and Belgium, Czech Republic, France, Germany, Greece, Italy, Norway, and Spain under grant agreement no. 101093441 (SPACE).
\end{acknowledgements}
\bibliographystyle{aa}
\bibliography{References}

\clearpage           

\begin{appendix}

\section{Turbulence driving mechanisms }\label{app:driving}
In this appendix we provide a brief description of the different mechanisms used to drive turbulence in the \texttt{FRIGG} simulations. Section~\ref{sn_driving} describes how turbulence is driven by SN feedback in the \texttt{SN} runs, while Sect.~\ref{turb_forcing} explains how turbulence is driven by large-scale forcing in the \texttt{TURB} runs.

\subsection{Supernova driving}\label{sn_driving}

In this work, the \texttt{SN} runs adopt a time-dependent UV background, and their numerical suffix indicates the initial mid–plane density (in cm\(^{-3}\)). 
Two members of this series, \texttt{SN-1} and \texttt{SN-0.66}, were presented in \citet{Colman_2025} under the names \texttt{n1\_rms00000} and \texttt{n0.66\_rms00000}, respectively.

In these runs sink particles \citep{bate_1995,federrath_2010_sink} are introduced when the density exceeds a threshold of \(10^{3}\,\mathrm{cm^{-3}}\). These particles can accrete mass from the surrounding gas. Each time sink particles accumulate $120\, \mathrm{M_\odot}$ , a stellar particle is created with a random mass between $8$ and $120\, \mathrm{M_\odot}$ drawn from a Salpeter initial mass function \citep{salpeter_1955}. The stellar particle is then assigned a main-sequence lifetime related to its mass. Once this lifetime ends, a SN explosion is triggered, injecting  $10^{51}\, \mathrm{erg}$ of thermal energy and $4 \times 10^{43}\, \mathrm{g\,cm\, s^{-1}}$ of radial momentum  into the surrounding cells \citep{iffrig_2015}. To avoid very short time steps, a temperature limit ($2\times10^6\, 
\mathrm{K}$) and a velocity limit ($200\, \mathrm{km\, s^{-1}}$) are set. 

\subsection{Turbulent forcing}\label{turb_forcing}
In the \texttt{TURB} runs, turbulence is injected following the Ornstein–Uhlenbeck (OU) process \citep{Eswaran_1988,Schmidt_2006,Schmidt_2009,Federrath_2010} as described in \citet{brucy_2023}.  
At each time step ($dt \simeq 0.4~\mathrm{Myr}$), a two–dimensional random acceleration field is  evolved in Fourier space,and then transformed back into real space.
The forcing is confined to the disk plane and restricted to large-scale Fourier modes with wavelengths between $L_{\rm box}/3$ and $L_{\rm box}$, with the maximum weight assigned to the mode corresponding to a wavelength of $L_{\rm box}/2$.
This mimics the injection of kinetic energy from processes at the galactic scale.

We adopt a compressive fraction $\chi = 0.5$ (the ratio of compressive to solenoidal modes contributions), which corresponds to the natural mixture of one‑third compressive and two‑thirds solenoidal modes projection \citep{Schmidt_2009, Federrath_2010}. The parameter $f_{\mathrm{rms}}$ controls the power injected by the forcing into the simulation; its value varies in the four runs listed above.
Each simulation was evolved for at least two turbulent crossing times, ensuring that the injected turbulence has fully developed throughout the box.

\section{Synthetic and observational sky maps}\label{app:cake_plots}

\begin{figure*}[t]
\centering
\begin{subfigure}[t]{0.33\textwidth}
  \centering
  \includegraphics[width=\linewidth]{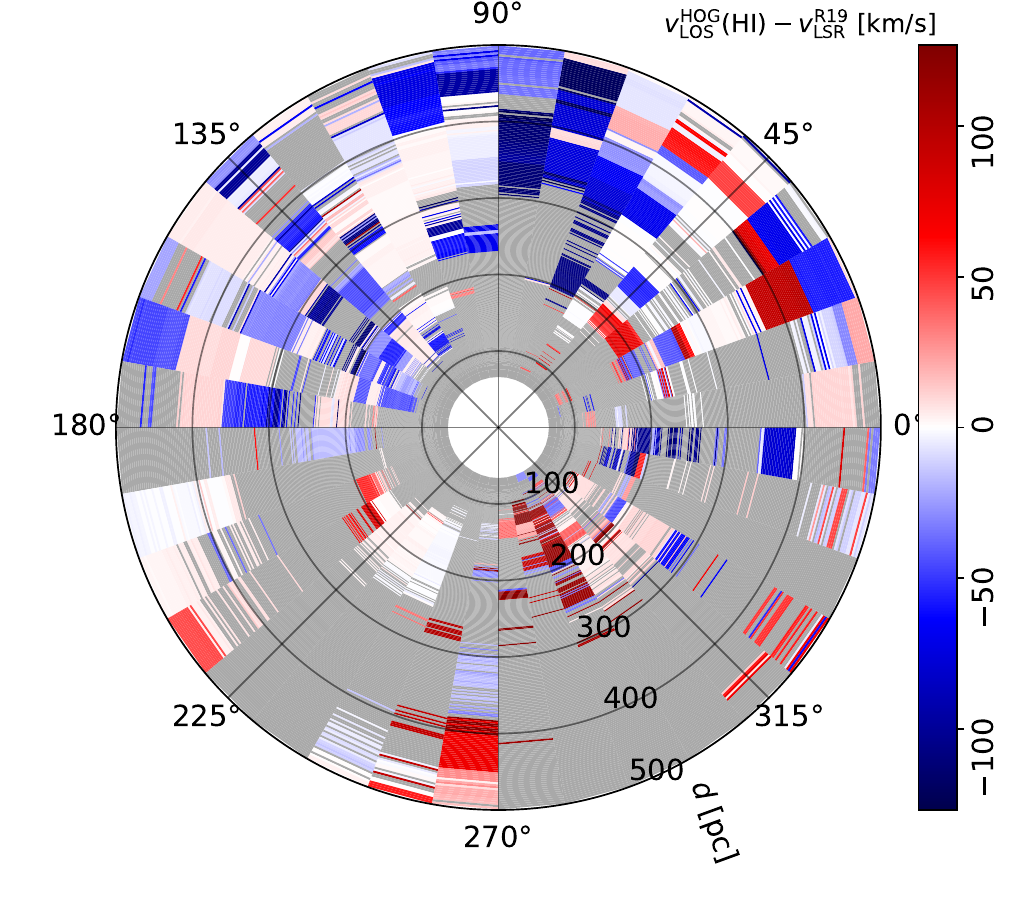}
  \subcaption{}
\end{subfigure}\hspace{2cm}
\begin{subfigure}[t]{0.33\textwidth}
  \centering
  \includegraphics[width=\linewidth]{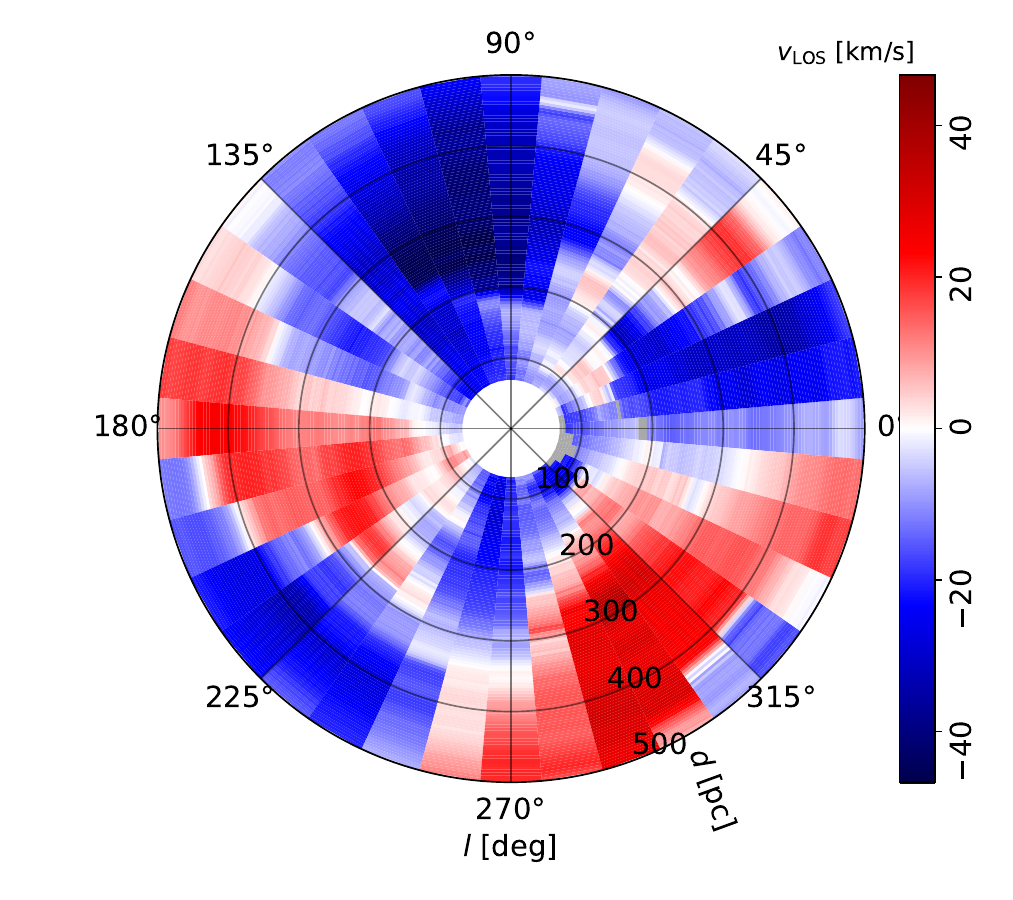}
  \subcaption{}
\end{subfigure}

\caption{Observed and simulated LOS velocity maps before applying the observational mask. Panel (a): Observed \ion{H}{i} LOS velocity map in \citetalias{Soler_2025}. Panel (b): Example of one full synthetic LOS velocity map from \texttt{TURB-15}.}
\label{fig:cake_full}
\end{figure*}

\begin{figure*}[t]
\centering
\begin{subfigure}[t]{0.33\textwidth}
  \centering
  \includegraphics[width=\linewidth]{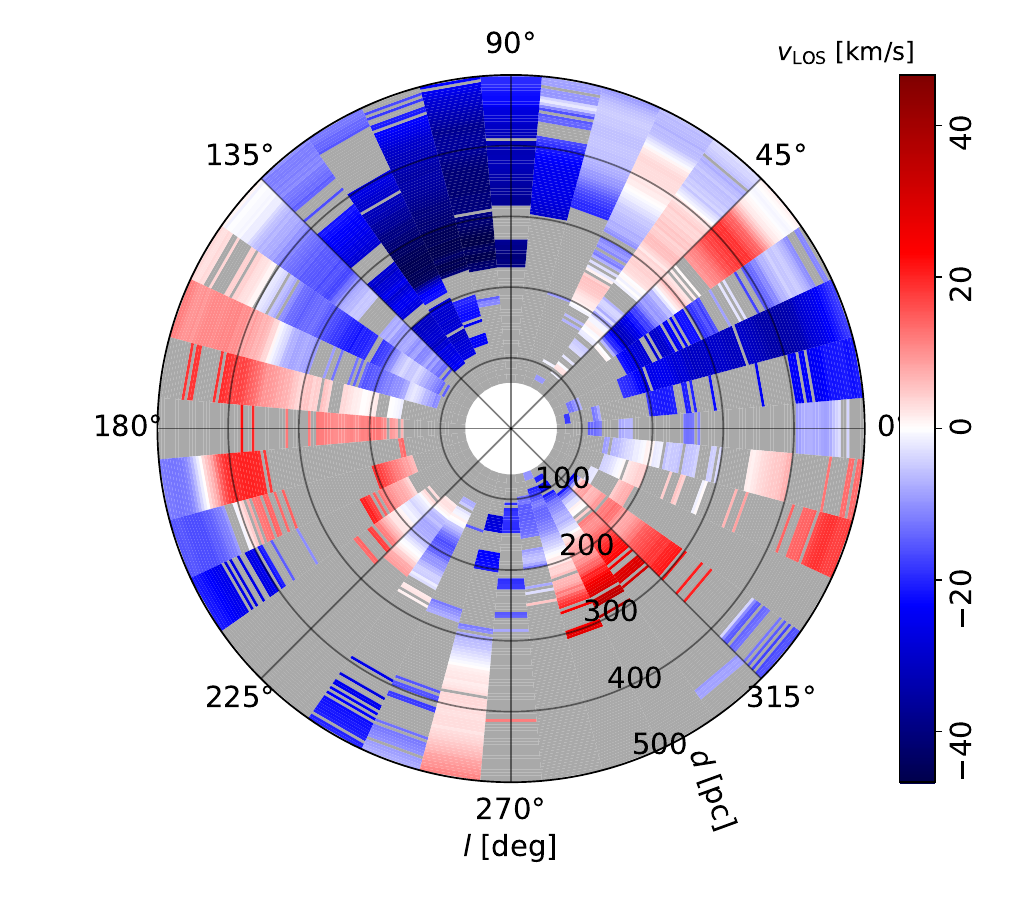}
  \subcaption{}
\end{subfigure}
\hspace{2cm}
\begin{subfigure}[t]{0.33\textwidth}
  \centering
  \includegraphics[width=\linewidth]{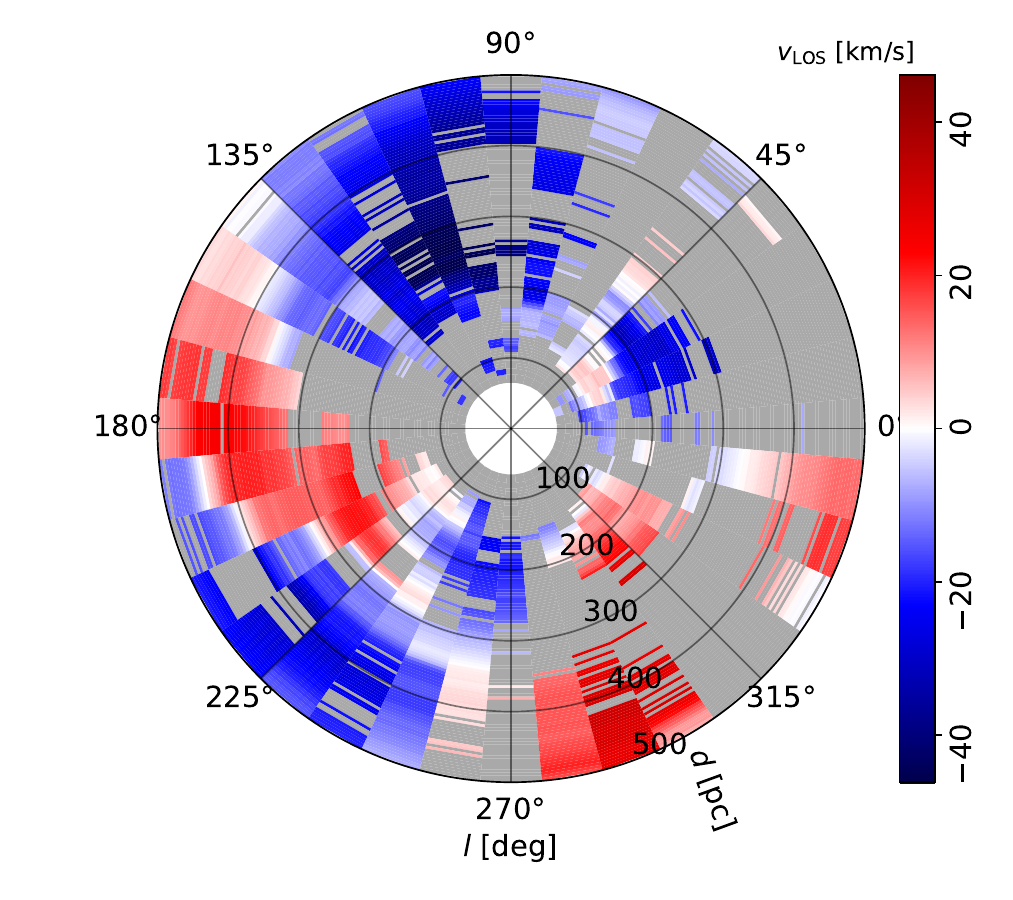}
  \subcaption{}
\end{subfigure}

\vspace{0.5cm}

\begin{subfigure}[t]{0.33\textwidth}
  \centering
  \includegraphics[width=\linewidth]{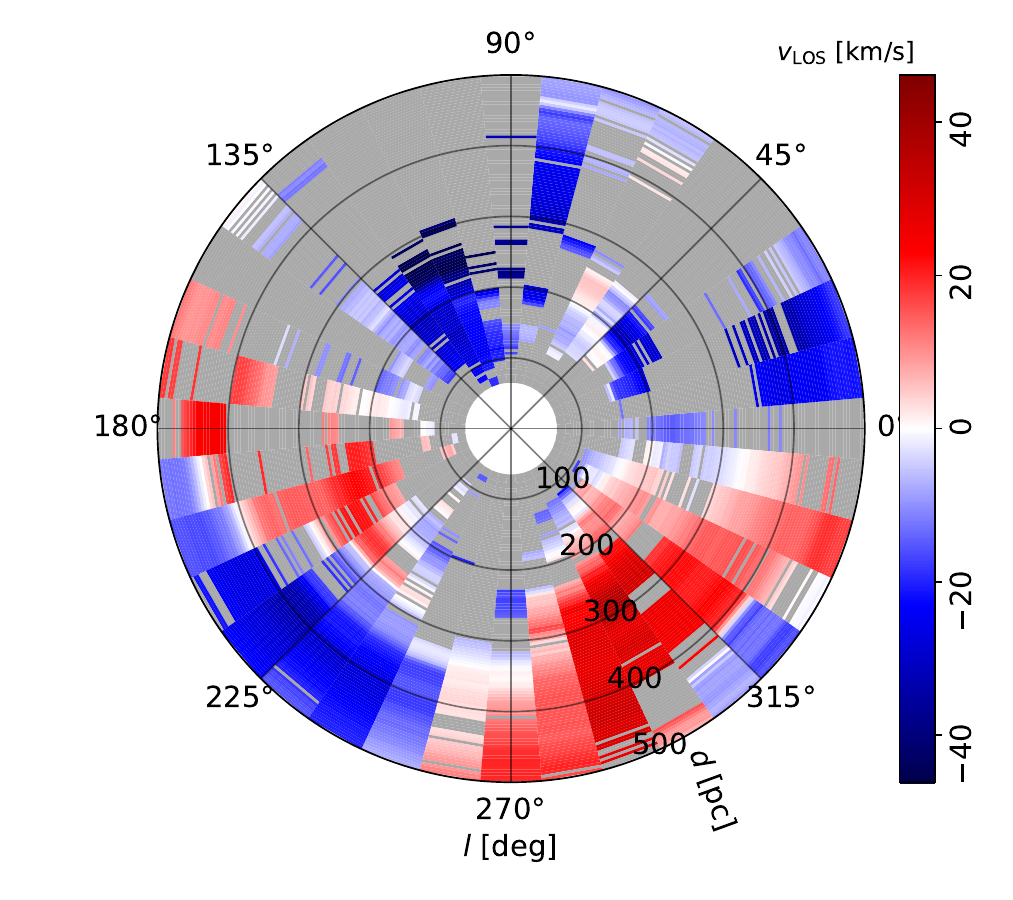}
  \subcaption{}
\end{subfigure}
\hspace{2cm}
\begin{subfigure}[t]{0.33\textwidth}
  \centering
  \includegraphics[width=\linewidth]{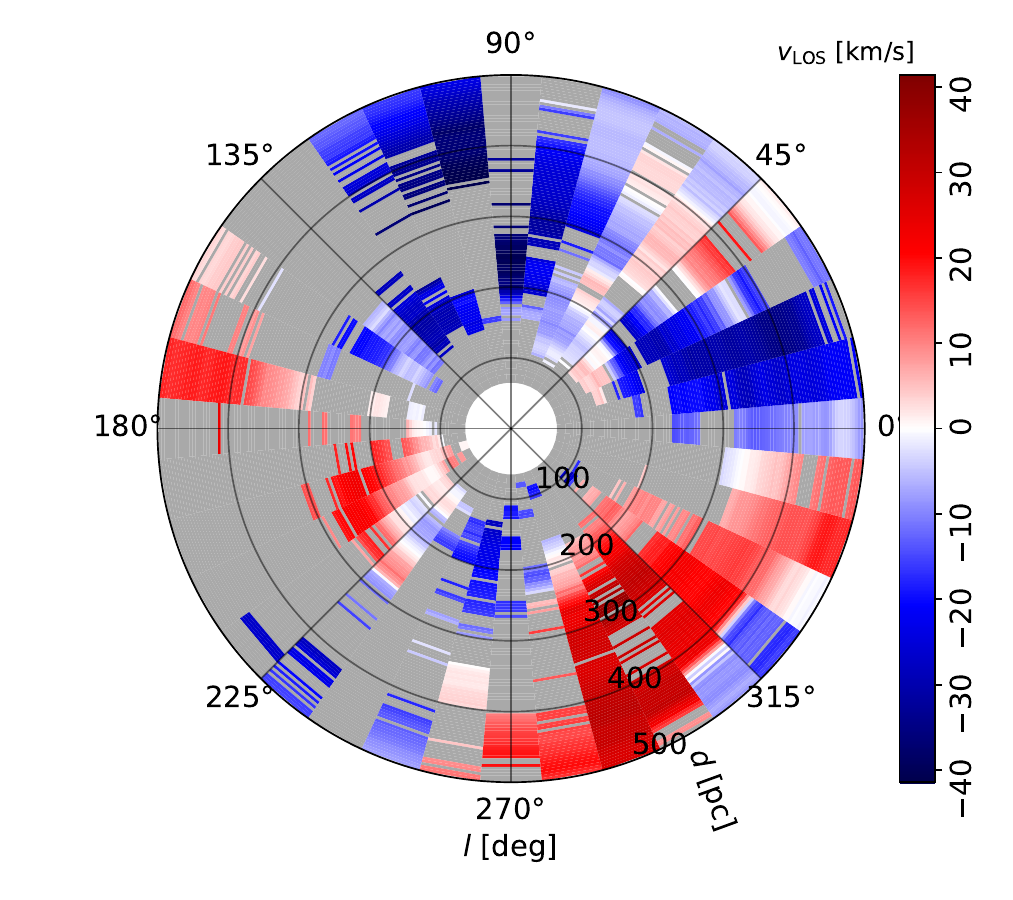}
  \subcaption{}
\end{subfigure}

\caption{Masked synthetic sky maps obtained by applying the observational sampling pattern at different rotation angles to the full synthetic map shown in Fig.~\ref{fig:cake_full}\,(b): Rotation \(0^\circ\) (panel a),  \(90^\circ\) (b), \(180^\circ\) (c), and \(270^\circ\) (d).}

\label{fig:cake_masked}
\end{figure*}

In this appendix we present a visual comparison between the
\ion{H}{i} sky map from \citetalias{Soler_2025} and an example of the synthetic sky maps generated from the simulations.  
Figure~\ref{fig:cake_full} shows side–by–side the observational map and a full
unmasked synthetic map from the \texttt{TURB-15} run.  
Figure~\ref{fig:cake_masked} then shows the four masked realizations
obtained by applying the observational sampling pattern after rotations of
\(0^\circ\), \(90^\circ\), \(180^\circ\), and \(270^\circ\) in longitude.
These masked maps are the ones used in the velocity–dispersion analysis.

\section{Evolution of the velocity PDF}\label{app:pdf}

In this appendix we show how the probability density function (PDF) of
the line--of--sight velocity evolves purely as a consequence of applying
successive numerical processing steps to the simulations.

\begin{figure*}[t]
    \centering
    \includegraphics[width=\linewidth]{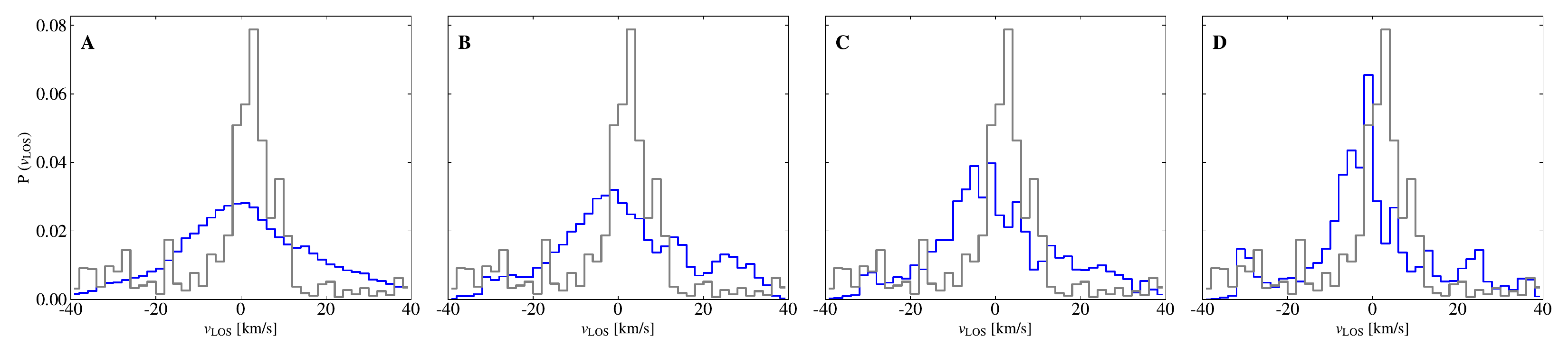}
    \caption{Evolution of the velocity PDF as successive numerical processing
    steps are applied to one \texttt{TURB-15} map.  
    The gray histogram shows the \ion{H}{i} PDF from \citetalias{Soler_2025}
    ($70$--$500$~pc, $|v_{\rm los}| \le 40~\mathrm{km\,s^{-1}}$).  
    Blue histograms correspond to (A) the intrinsic full-resolution 3D
    simulation, (B) the 2D \((x,y)\) map obtained by averaging along \(z\),
    (C) the synthetic sky projection (4~pc radial, \(10^\circ\) angular
    sampling), and (D) the same projection after applying one observational
    mask.}
    \label{fig:pdf_evolution}
\end{figure*}

Figure~\ref{fig:pdf_evolution} shows four panels comparing the \ion{H}{i}
velocity PDF from \citetalias{Soler_2025} (gray histogram; $70$--$500$~pc,
$|v_{\rm los}| \le 40~\mathrm{km\,s^{-1}}$) with the corresponding
mass--weighted distribution obtained from one \texttt{TURB-15} synthetic map
(blue), analyzed under four different configurations:

\begin{itemize}
    \item \textbf{Panel a:} Point spread function of $v_{los}$ for all simulation cells satisfying
    $70 \le r \le 500$~pc and $|z| \le r \sin 5^\circ$ in the full 4~pc
    uniform cube ($\sim 6\times10^{5}$ cells).  

    \item \textbf{Panel b:} Same spatial selection, but after collapsing
    the cube into a 2D \((x,y)\) map by averaging over all cells along the
    entire vertical LOS (all \(z\) at fixed \(x,y\); $\sim 5\times10^{4}$ cells).

    \item \textbf{Panel c:} Cells from the synthetic sky projection with
    4~pc radial resolution and $10^\circ$ angular sampling  
    ($\sim 3.8\times10^{3}$ cells).  

    \item \textbf{Panel d:} Same as panel c, but after applying one of the
    observational masks (Appendix~\ref{app:cake_plots}), leaving
    $\sim 1.2\times10^{3}$ sampled cells.  
   
\end{itemize}

A clear evolution is visible from panels a to d: the intrinsic simulation PDF
is smooth, but becomes progressively narrower and more irregular as averaging,
angular binning, and sky incompleteness are introduced.  
This shows that features such as a narrow central peak and long
high--velocity tails -- characteristic of the observational PDF -- can arise from
numerical sampling effects alone.  
We do not claim that every synthetic sky reproduces the observed PDF, but the
example shown here illustrates how limited resolution and partial coverage can
produce these features.

\section{Dependences on the velocity threshold}\label{app:diff_vel_thresholds}

\begin{figure}[t]
  \centering
  \includegraphics[width=\linewidth]{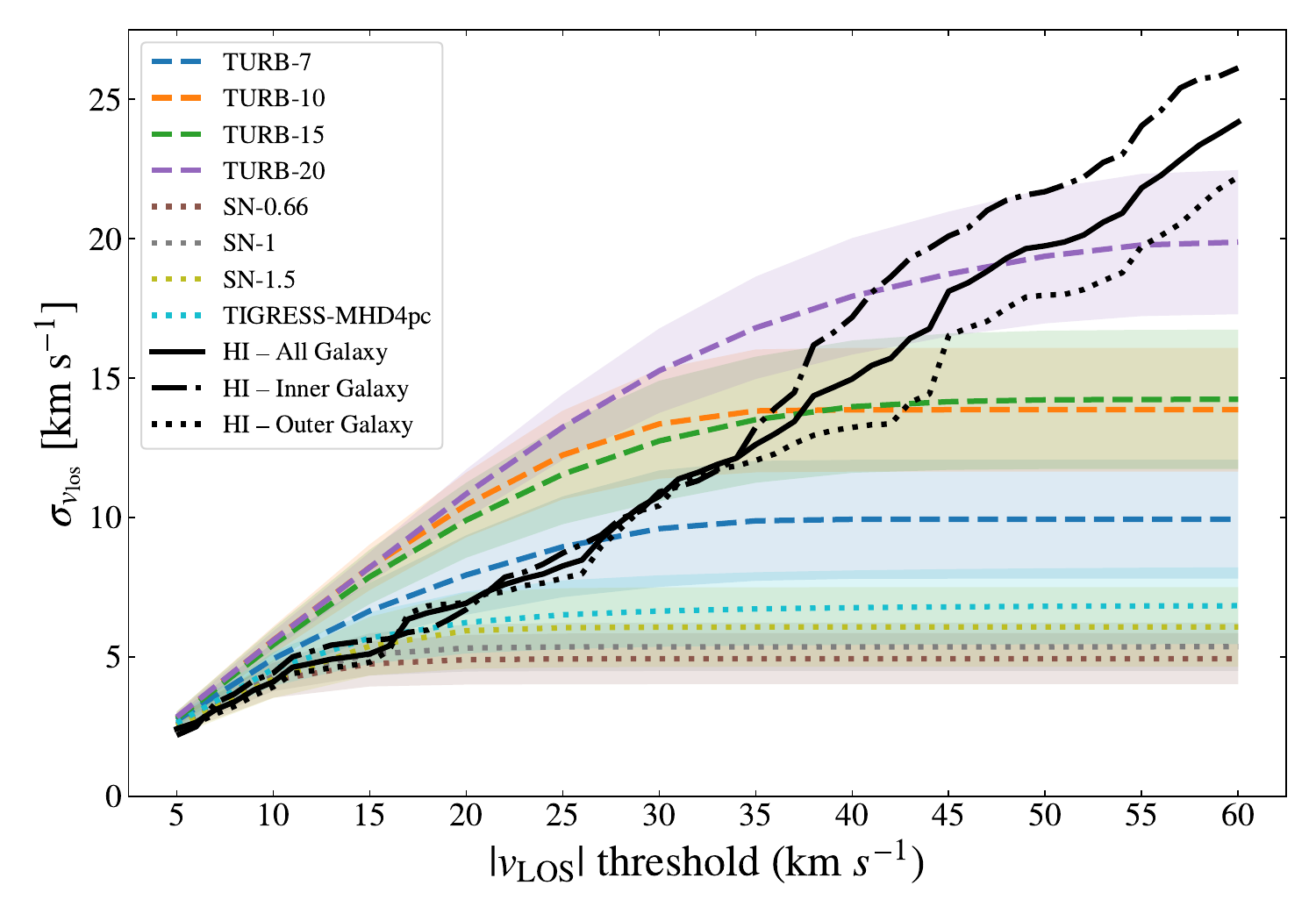}
  \caption{Evolution of the mass-weighted LOS velocity dispersion as a function of the imposed LOS–velocity threshold for both simulations and observations. Colored curves show the simulated models, averaged over the corresponding sets of synthetic sky maps, with SN-driven runs shown as colored dotted lines and large-scale turbulent-forcing runs as colored dashed lines. Black curves correspond to the observational \ion{H}{i} data from \citetalias{Soler_2025}: solid for the full dataset, dotted for the outer Galaxy, and dash-dotted for the inner Galaxy.}
  \label{fig:evolutin_thresholds}
\end{figure}

In Sect.~\ref{sec:vel_disp_comp}, where only velocities up to  $30~\mathrm{km\,s^{-1}}$ were considered, we showed that SN-driven models yield velocity dispersions below the \ion{H}{i} value reported by \citetalias{Soler_2025}, implying that an additional large-scale driving mechanism is required. The observed values lie between those of the \texttt{TURB-7} and \texttt{TURB-10} simulations.

Appendix~\ref{app:pdf} further demonstrates that the non-smooth shape of the observed \ion{H}{i} LOS velocity PDF, characterized by a sharp central peak and extended high-velocity tails, can be recovered from the simulations once the observational sampling pattern and resolution are applied. This indicates that part of the observed high-velocity structure may arise from resolution and completeness effects and may reflect the way velocities in these ranges are sampled rather than actual physical anomalies.

To estimate how much large-scale forcing would be required if velocities above \(30~\mathrm{km\,s^{-1}}\) are taken into consideration, we examine how the velocity dispersion varies when imposing different LOS-velocity thresholds, applied to both simulations and observations. Figure~\ref{fig:evolutin_thresholds} shows the evolution of the average mass-weighted velocity dispersion as a function of the imposed threshold. All simulations display a similar behavior: the dispersion increases with the threshold and eventually saturates once the high-velocity tail is fully recovered. Supernova-driven runs saturate at a threshold of $\sim 20$--$25~\mathrm{km\,s^{-1}}$, whereas the large-scale-forced runs saturate at significantly higher values, from $\sim 30$--$35~\mathrm{km\,s^{-1}}$ for \texttt{TURB-7} up to $\sim 55~\mathrm{km\,s^{-1}}$ for \texttt{TURB-20}.

For the observational data, we also separate the inner and outer Galaxy sight lines. Up to $\sim 35~\mathrm{km\,s^{-1}}$, both regions present similar behavior, then they diverge after that, with the inner Galaxy showing larger dispersions. 
At higher velocity thresholds, the continued increase of the observed velocity
dispersion is likely dominated by contamination from noise and contributions
from distant, nonlocal high-velocity gas.
When comparing simulations with the full Galaxy observational curve, the closest match depends on the chosen threshold: at $30~\mathrm{km\,s^{-1}}$ \texttt{TURB-7} matches best; at $40~\mathrm{km\,s^{-1}}$ the closest model becomes \texttt{TURB-10} or \texttt{TURB-15}; and at $50~\mathrm{km\,s^{-1}}$, \texttt{TURB-20} provides the best agreement. However, reliably determining the forcing amplitude required at these higher thresholds will ultimately require more complete sky coverage and more confident \ion{H}{i} measurements at high velocities, in order to distinguish genuine turbulent signal from distant motions or instrumental contamination.

\section{Dependences on the Galactic rotation model}\label{app:galacticrotation}

In this appendix we consider the effects introduced by the uncertainties in the Galactic rotation models employed to subtract Galactic rotation from kinematic tomography LOS velocity estimates in \citetalias{Soler_2025}.
We focus on the consequences of two aspects of the Galactic rotation model.
First, we tested the choice of Galactic rotation model by considering alternatives.
Second, we evaluated the effect of uncertainties in the parameters of the Galactic rotation model in \cite{Reid_2019}.

We considered three alternatives to the \cite{Reid_2019} model: a flat rotation curve and the models in \cite{brand_1993} and \cite{Reid_2014}.
The \cite{Reid_2014} model is the previous iteration of the \cite{Reid_2019}, obtained with 50\% fewer parallax measurements.
The \cite{brand_1993} model is a landmark reference, based on spectrophotometric distances to H{\sc ii} regions and H{\sc i} 21-cm emission observations.
Finally, the flat rotation curve, constant circular speed with Galactocentric radius, is an acceptable approximation for the Milky Way’s disk dynamics around the solar radius.

\begin{figure*}[t]
    \centering{
    \includegraphics[width=0.33\linewidth]{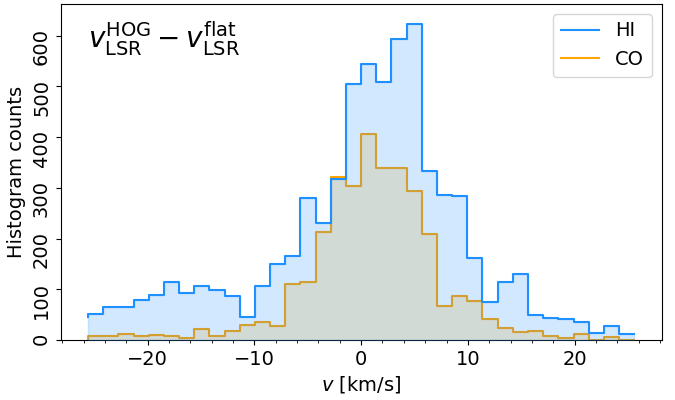}
    \includegraphics[width=0.33\linewidth]{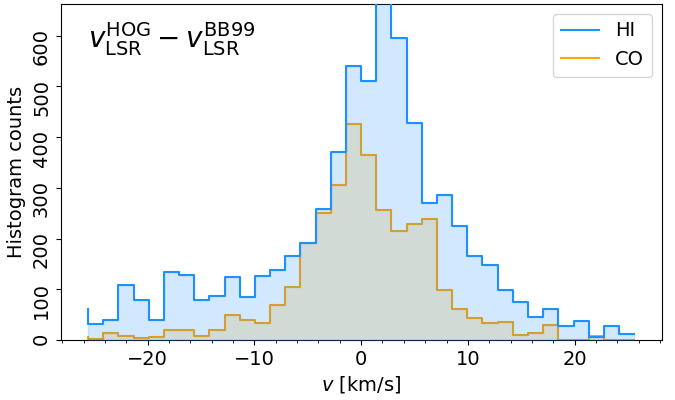}
    \includegraphics[width=0.33\linewidth]{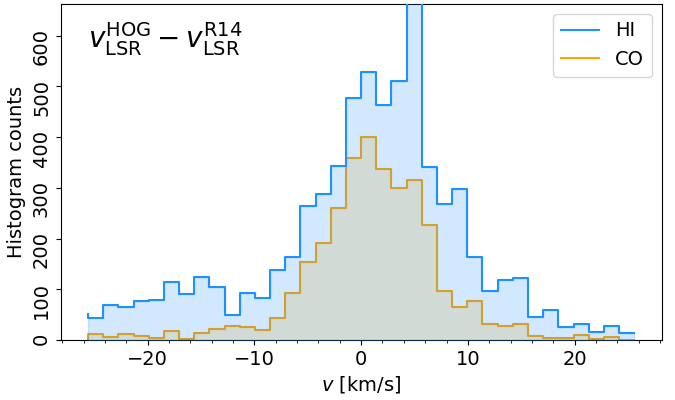}
    }
    \caption{Same as Fig. 12 in \citetalias{Soler_2025} but for three different Galactic rotation models.}
   \label{fig:sigmavModels}
\end{figure*}

Table~\ref{table:sigmavMultiModel} presents the standard deviations of the LOS velocities obtained after subtracting the different models of Galactic rotation from the kinematic tomography results in \citetalias{Soler_2025}.
We consider two distance ranges in the input data: 70\,$<$\,$d$\,$<$\,500\,pc, which matches the size of the simulation domain, and 70\,$<$\,$d$\,$<$\,1250\,pc, which encompasses the distance range in the \citetalias{Soler_2025} reconstruction.
The results reported in Table~\ref{table:sigmavMultiModel} demonstrate that there is very little variation in the resulting standard deviation of noncircular motions, as expected from the nearly constant circular speed around the solar radius.
The resulting histograms of residual motions, shown in Fig.~\ref{fig:sigmavModels}, indicate that the global distributions are very similar for the different Galactic rotation models.

\begin{table}[ht!]
\caption{Estimated standard deviation of residual velocities ($\sigma_{v}$) after subtraction of different Galactic rotation models.}\label{table:sigmavMultiModel}
\centering
\begin{tabular}{lcccc}
\hline\hline
  & \multicolumn{2}{c}{70\,$<$\,$d$\,$<$\,500\,pc} & \multicolumn{2}{c}{70\,$<$\,$d$\,$<$\,1250\,pc}\\
Galactic rotation model & $\sigma^{\rm HI}_{v}$ & $\sigma^{\rm CO}_{v}$ & $\sigma^{\rm HI}_{v}$ & $\sigma^{\rm CO}_{v}$ \\
 & [km/s] & [km/s] & [km/s] & [km/s] \\
\hline
\cite{Reid_2019}  & 10.8 & 6.0 & 10.8 & 6.6 \\ 
\cite{Reid_2014}  & 10.6 & 5.9 & 10.5 & 6.3 \\
\cite{brand_1993} & 10.7 & 6.1 & 10.7 & 6.6\\ 
Flat              & 10.6 & 6.0 & 10.4 & 6.3 \\
\hline
\end{tabular}
\end{table}

We also tested the effects of the uncertainties in \cite{Reid_2019} model parameters.
For that purpose, we produce $N_{\rm MC}$\,$=$\,1000 realizations of the rotation curve by Monte Carlo sampling its fitted parameters: the Sun distance from the Galactic center ($R_{0}$), the three-dimensional motion of the Sun in its orbit about the Galaxy ($U_{\odot}$, $V_{\odot}$, $W_{\odot}$), the average peculiar motion of the stars ($\bar{U_{\sc s}}$, $\bar{V_{\sc s}}$), and the universal rotation curve parameters ($a_{2}$,$a_{3}$).
We sampled the distribution of these parameters using the central values and standard deviations for the A1 model reported in table~3 of \citep{Reid_2019} using the {\tt resample\_params} routine in the {\tt kd} package \citep{wegner2018}.
Table~\ref{able:sigmavR19sampling} presents the resulting mean, median, and standard deviations obtained by subtracting each of the $N_{\rm MC}$ realizations of the circular motions from the kinematic tomography results in \citetalias{Soler_2025}.

Table~\ref{able:sigmavR19sampling} presents statistical parameters that characterize the $\sigma_v$ distribution for $N_{\rm MC}$ realizations of the \cite{Reid_2019} model. 
Given that $\sigma_{v}$ is a quadratic quantity, it is expected that the mean values are positively biased and the statistical trend is better by the median, which is within 10\% of the values reported in \citetalias{Soler_2025}.
The resulting $\sigma_{v}$ standard deviation is also below 10\% of the mean and median values of that quantity.
Thus, we conclude that the uncertainties in the \cite{Reid_2019} model do not significantly challenge the conclusions in \citetalias{Soler_2025}.

The functional form of the \cite{Reid_2019} model implies that $\sigma_{v}$ is unlikely to follow a Gaussian distribution for variations in the input parameters.
Thus, we also report the 1st, 25th, 75th, and  99th percentiles of the $\sigma_{v}$ distribution obtained from Monte Carlo sampling.
Their values indicate that the distribution is right-skewed, that is, the statistical mode is below the median and the mean.
The reported percentiles also indicate that variations in the \cite{Reid_2019} model are unlikely to produce significant changes below the median value, 10.5\,km/s for \ion{H}{i}.

\begin{figure*}[t]
    \centering{
    \includegraphics[width=0.33\linewidth]{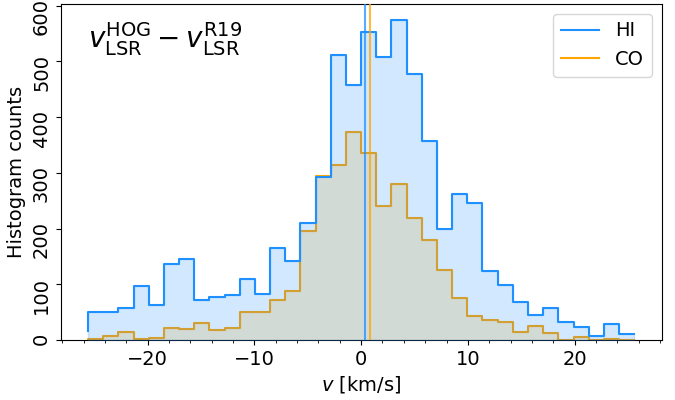}
    \includegraphics[width=0.33\linewidth]{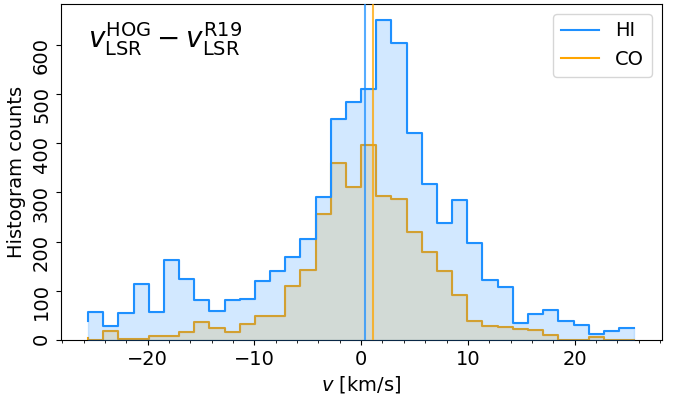}
    \includegraphics[width=0.33\linewidth]{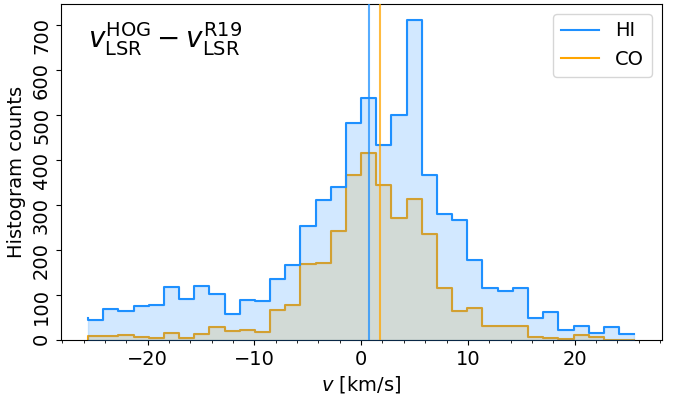}
    }
    \centering{
    \includegraphics[width=0.33\linewidth]{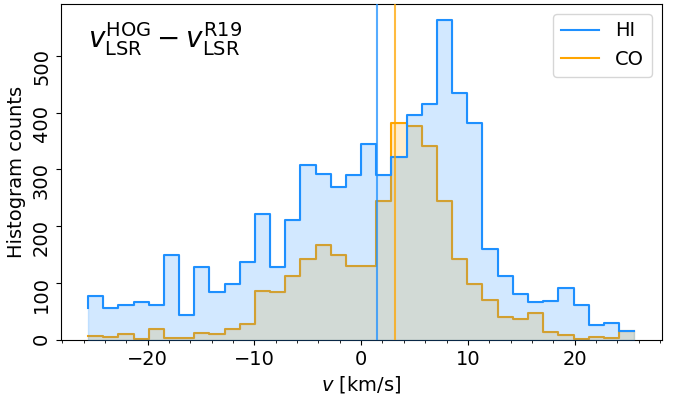}
    \includegraphics[width=0.33\linewidth]{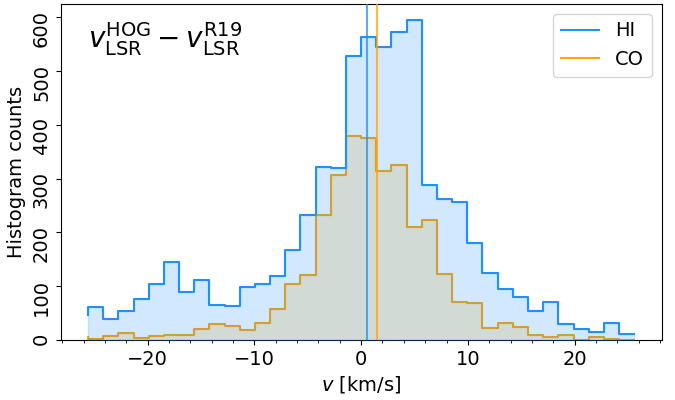}
    \includegraphics[width=0.33\linewidth]{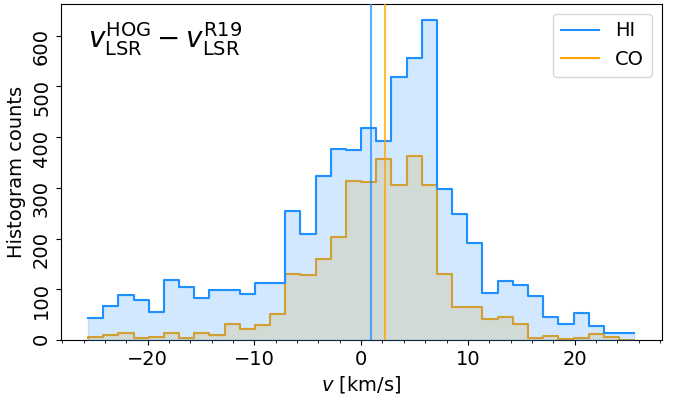}
    }
    \caption{Same as Fig. 12 in \citetalias{Soler_2025} but for the sampling of parameters in the \citep{Reid_2019} Galactic rotation model.}
    \label{fig:sigmavMC}
\end{figure*}

Figure~\ref{fig:sigmavMC} illustrates the variation in the histograms of residual velocities for a selection of random realizations of the \cite{Reid_2019} model.
The differences in the histograms show that specific features in the distribution can be introduced by the variations in the subtracted Galactic rotation model.
However, the convergence of the mean and median $\sigma_{v}$ values indicates that these excursions are not statistically significant relative to the results of numerical simulations.

\begin{table}[ht!]
\caption{Statistics of standard deviation of residual velocities ($\sigma_{v}$) for the Monte Carlo sampling of 1000 combinations of the \cite{Reid_2019} Galactic rotation models.}\label{able:sigmavR19sampling}
\centering
\begin{tabular}{lcccc}
\hline\hline
  & \multicolumn{2}{c}{70\,$<$\,$d$\,$<$\,500\,pc} & \multicolumn{2}{c}{70\,$<$\,$d$\,$<$\,1250\,pc}\\
Galactic rotation model & H{\sc i} & CO & H{\sc i} & CO \\
 & [km/s] & [km/s] & [km/s] & [km/s] \\
\hline
Mean & 11.6 & 6.9& 11.5 & 7.6  \\ 
Median & 11.1 & 6.4 & 10.9 & 6.9 \\
Standard deviation & 1.2 & 1.2 & 1.3 & 1.6\\ 
1st percentile & 10.6 & 5.9 & 10.5 & 6.3\\ 
25th percentile & 10.8 & 6.1 & 10.7 & 6.5\\ 
75th percentile & 12.0 & 7.4 & 12.3 & 8.6\\ 
99th percentile & 16.0 & 10.9 & 16.3 & 13.1\\ 
\hline
\end{tabular}
\end{table}

\section{Dependence on temperature thresholds}\label{app:diff_cuts_temp}

\begin{figure*}[t]
    \centering
    \includegraphics[width=0.33\linewidth]{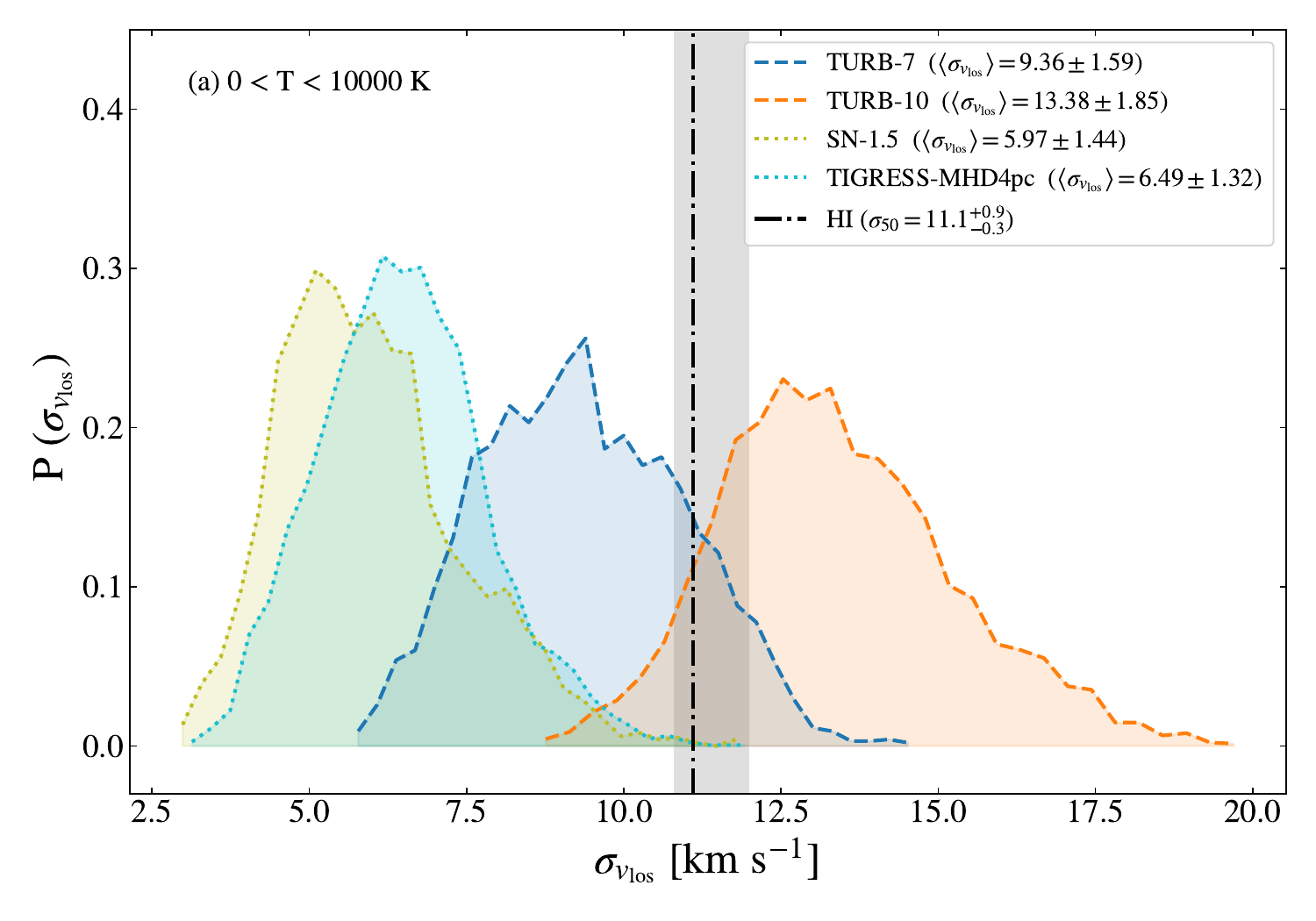}
    \includegraphics[width=0.33\linewidth]{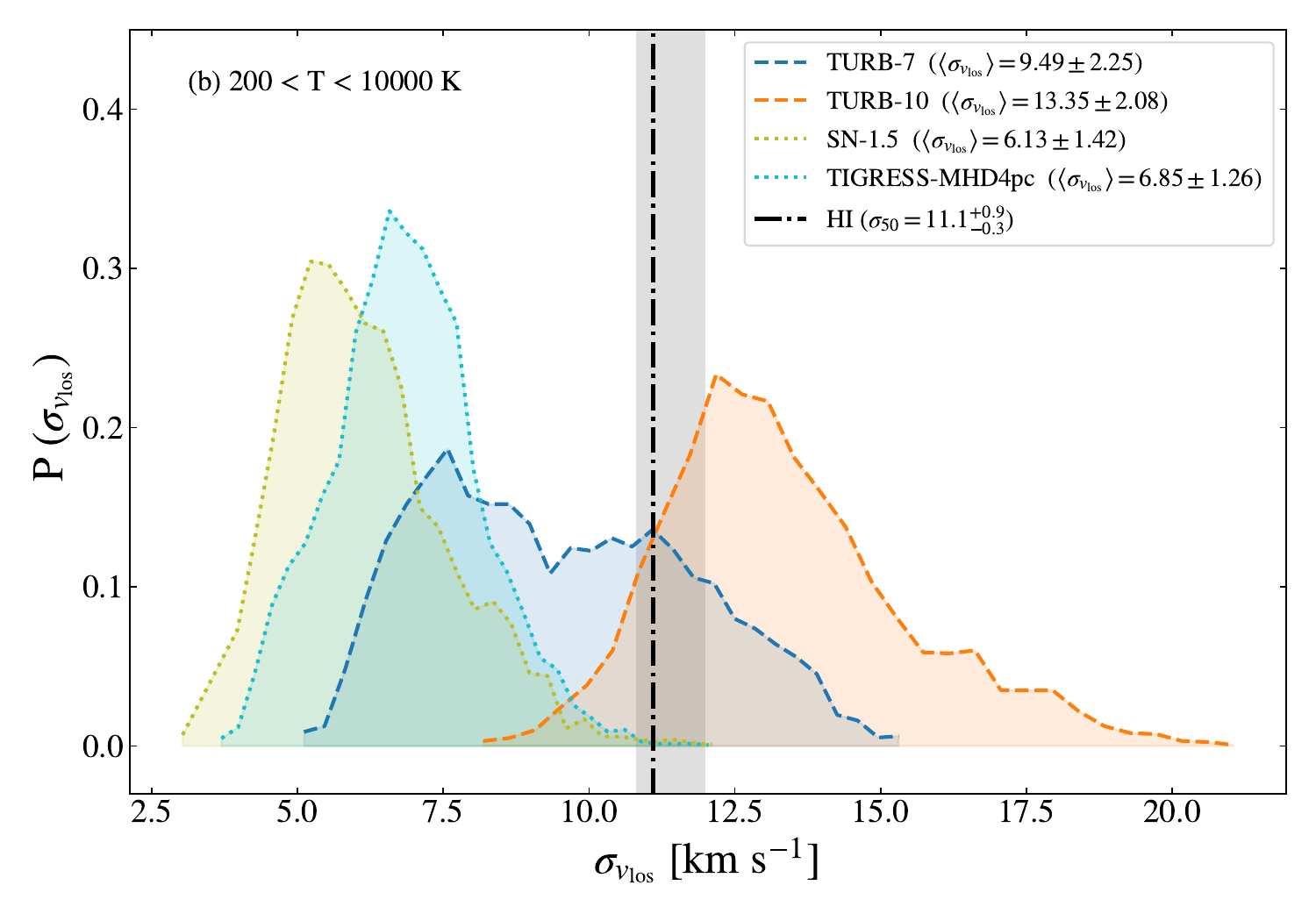}
    \includegraphics[width=0.33\linewidth]{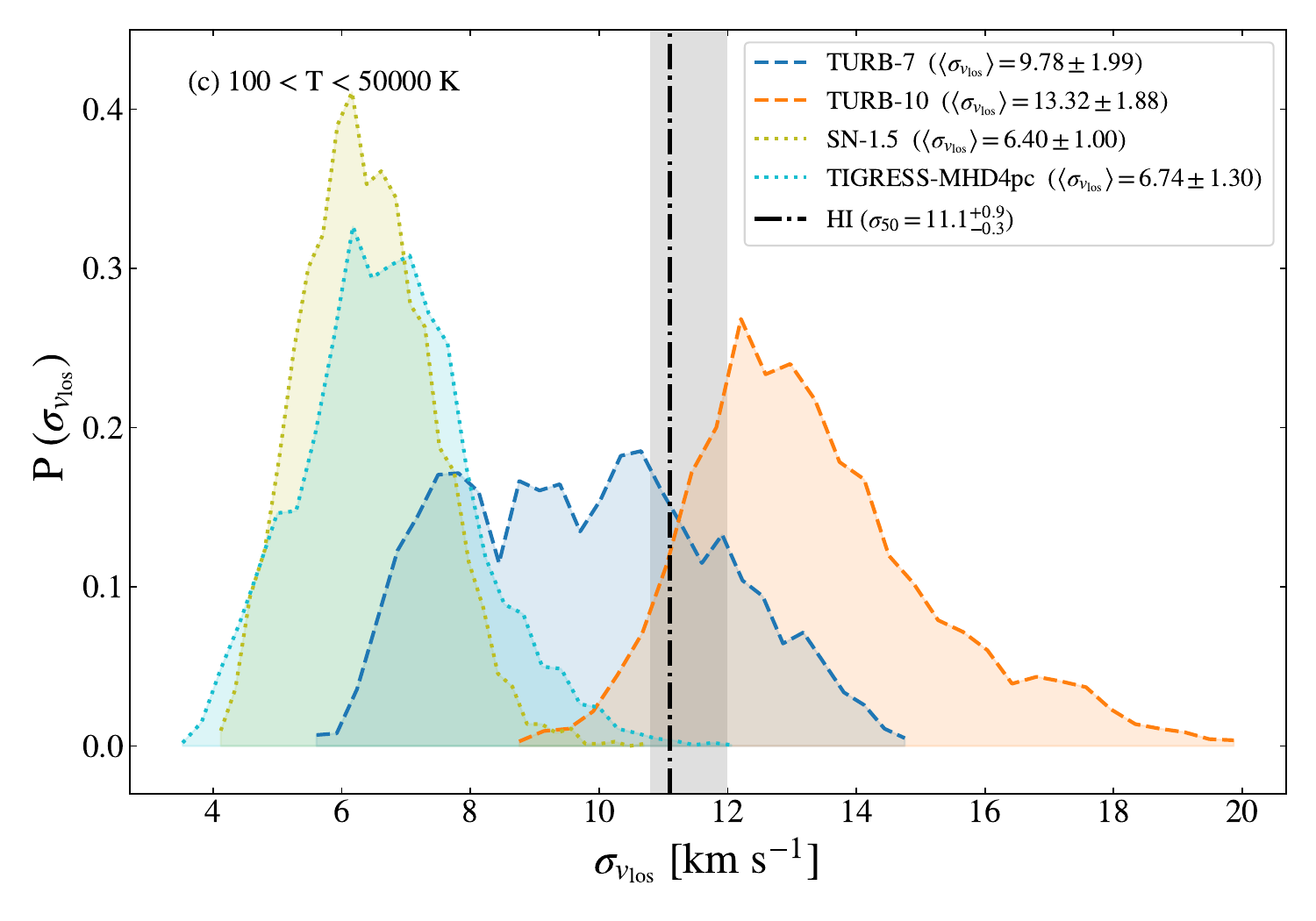}

    \caption{Same as Fig.~\ref{fig:sigma_distribution} but selecting gas within different temperature ranges: 
    (a) $0 < T < 10000\,\mathrm{K}$;
    (b) $200 < T < 10000\,\mathrm{K}$;
    (c) $100 < T < 50000\,\mathrm{K}$.}
   \label{fig:sigma_diff_temp}
\end{figure*}

In this appendix we investigate how the inferred velocity dispersions depend on the adopted temperature thresholds used to select the \ion{H}{i} component in the simulation.Observationally, neutral atomic hydrogen spans a broad temperature range. The cold neutral medium (CNM) typically occupies $T \sim 25$–$250\,\mathrm{K}$, the unstable neutral medium (UNM) $T \sim 250$–$4000\,\mathrm{K}$, and the warm neutral medium (WNM) $T \sim 6000$–$8000\,\mathrm{K}$  \citep{McClure-Griffiths_2023}. 
In the \texttt{FRIGG SN} runs, \cite{Colman_2025} classify the ISM gas phases using temperature thresholds, with the atomic phases occupying temperatures below $\sim 10000\,\mathrm{K}$. 
In the TIGRESS simulation \citep{tigress_2017}, gas phases are defined using temperature thresholds, with neutral gas below $20000\,\mathrm{K}$ and ionized gas above this value.
Although molecular hydrogen is present at the lowest temperatures, near those of the CNM, and ionized gas appears at temperatures approaching and exceeding those of the WNM, the total mass over intermediate temperatures is dominated by neutral atomic hydrogen. 
In their Solar Cirle simulation, \citet{Tress_2025} shows that neutral atomic gas dominates the mass distribution particularly between $\sim 100$ and $10000\,\mathrm{K}$ where it exceeds both molecular (H$_2$) and ionized (H$^+$) gas. At lower temperatures ($\lesssim 100\,\mathrm{K}$), the mass is dominated by molecular gas, while at higher temperatures ($\gtrsim 20000,\mathrm{K}$), the ionized gas becomes the dominant component.

In the main text, we adopt a temperature threshold of $100 < T < 10000\,\mathrm{K}$ to approximate the neutral atomic hydrogen component.  To assess the robustness of this choice, we recompute the LOS velocity dispersions using three alternative temperature ranges:  
(1) $0 < T < 10000\,\mathrm{K}$, which includes the coldest neutral atomic gas but also the molecular one; 
(2) $200 < T < 10000\,\mathrm{K}$, which excludes most of the molecular gas but also removes part of the CNM;  
(3) $100 < T < 50000\,\mathrm{K}$, to evaluate the impact of including a fraction of the ionized gas.

Figure~\ref{fig:sigma_diff_temp} shows the distributions of LOS velocity dispersions obtained for different temperature thresholds, for the simulations \texttt{SN-1.5}, \texttt{TIGRESS-MHD4pc}, \texttt{TURB-7}, and \texttt{TURB-10}. 
Varying the adopted temperature range produces only minor changes in the resulting dispersions. The relative order of the simulations remains unchanged, and the mean values shift by a few tenths of kilometers per second at most. In particular, SN-driven models remain systematically below the observational reference, even when molecular gas is excluded or a fraction of the ionized gas is included, while the large-scale forcing runs continue to exceed it. Our conclusions therefore do not depend sensitively on the adopted temperature range used to approximate the neutral atomic component.

\end{appendix}

\end{document}